# On some features of the solar proton event on 2021 October 28 – GLE73

**I. M. Chertok**[1]★

[1]*Pushkov Institute of Terrestrial Magnetism, Ionosphere and Radio Wave Propagation (IZMIRAN), Troitsk, Moscow 108840, Russia*

**ABSTRACT**

In addition to several recent articles devoted to the rare event of ground-level enhancement of the solar relativistic proton flux observed on 2021 October 28 – GLE73, we study the 10–100 MeV solar energetic particle (SEP) component of this event. Based on the GOES satellite data for 26 GLEs recorded since 1986, we have formed a scatter plot displaying the ratio of the peak fluxes of the >10 MeV ($J_{10}$) and >100 MeV ($J_{100}$) protons and their energy spectra. Two extreme characteristics of the prompt component of the SEP–GLE73 event were revealed: (1) very small $J_{10}$ and $J_{100}$ proton fluxes and (2) a very hard energetic spectrum in the 10–100 MeV range. There are only two events with these characteristics similar to SEP–GLE73 namely, GLE40 (1989 July 25) and GLE46 (1989 November 15). A correspondence was demonstrated between the hard frequency spectrum of microwave radio bursts of initiating flares and the hard SEP energy spectrum of these two and other GLEs. These results suggest that the flare magnetic reconnection both in the impulsive and post-eruption phases plays an important role in the acceleration of the SEP–GLE protons.

## 1 INTRODUCTION

The ascending phase of the solar Cycle 25 has been marked by two considerable, different in their nature and not quite ordinary space weather disturbances associated with solar and galactic cosmic rays. Both events were recorded by the world network of ground-based neutron monitors (NMs). These events took place within a week from 2021 October 28 to November 4, i.e. they were close in time. However, they occurred completely independently of each other and were initiated by eruptive flares in different active regions and by the corresponding coronal mass ejections (CMEs). Information about these events is available in the global NM database (NMDB; http://www.nmdb.eu), the Oulu database (http://cosmicrays.oulu.fi), the IZMIRAN database (ftp://cr0.izmiran.rssi.ru/COSRAY!/), as well as at the Web site of the Space Weather Prediction Centers (https://www.spaceweather.com), in the Preliminary Report and Forecast of Solar–Geophysical Data, No. 2409, 2410 (ftp://ftp.swpc.noaa.gov/pub/warehouse/2021/), and in other sources.

Firstly, we are talking about such a rare phenomenon as a ground level enhancement (GLE) of the solar relativistic proton count rate registered on October 28 by NMs around the world and designated as GLE73. That is how many events have been recorded since the beginning of observations in 1942 (see the GLE list at the NM databases indicated above).

★ E-mail: ichertok@izmiran.ru

The second disturbance occurred on 2021 November 3–4. It was a significant short-term density decrease of the galactic cosmic rays (GCR), the so-called Forbush decrease (FD) (see the reviews by Cane 2000 and Belov 2009). The peculiarity of this non-recurrent FD is that its magnitude was unusually large (up to 12%) for a fairly moderate peak magnetic field strength (~22 nT) and plasma velocity (~780 km/s) inside the interplanetary coronal mass ejection (ICME) observed in the near-Earth space by the Deep Space Climate Observatory (DSCOVR; https://www.spaceweather.com/images2021/05nov21/cmeimpact_data.jpg). Such a combination of parameters of this FD deserves a separate detailed study.

Here we focus on the analysis of the GLE73 event. In the first comprehensive article based on ground-based and space-borne observations of cosmic rays and solar activity, Papaioannou et al. (2022) analyzed in detail the main characteristics of GLE73, including such as its relation to a strong flare, broad halo CME and global extreme ultraviolet coronal wave, the solar release time of relativistic protons, their rigidity spectrum, anisotropy, and etc. The GLE73 is briefly compared with several previous GLEs that had similar solar drivers. Usually GLEs are high-energy component of the sufficiently intense solar energetic particle (SEPs) events registered by spacecraft. Hereinafter, the term SEP will refer to the proton flux in the energy range of 10–100 MeV, and the term GLE73, as before, to an enhancement of the relativistic proton flux.

This paper addresses the remarkably small flux and hard energy spectrum of the above-mentioned SEP and the corresponding characteristics of the flare microwave radio burst. In this context, the present paper should be considered as a kind of supplement to the article by Papaioannou et al. (2022). An overview of the SEP–GLE73 event is given in Section 2. Section 3 analyzes SEP in terms of the relationship between the fluxes and the energy spectrum of the >10 and >100 MeV protons typical for GLEs. The connection of these parameters with the frequency spectrum of flare microwave bursts is demonstrated in Section 4. The findings are summarized and discussed in Section 5.

## 2 OVERVIEW OF THE EVENT

Besides the aforementioned paper by Papaioannou et al. (2022), there are several other publications devoted to the analysis of the SEP–GLE73 event and associated solar activity phenomena: Battaglia, Collier, & Krucker (2022); Hou et al. (2022); Li et al. (2022); Klein et al. (2022); Li et al. (2022); Mishev et al. (2022); Xu et al. (2022). The solar manifestations of this event were observed with a number of spacecraft, in particular, with the Geostationary Operational Environmental Satellite (GOES; https://www.ngdc.noaa.gov/stp/satellite/goes/), the Solar and Heliospheric Observatory (SOHO; Domingo et al. 1995), the Solar Dynamics Observatory (SDO; Pesnell, Thompson, & Chamberlin 2012), the Solar TErrestrial RElations Observatory (STEREO-A; Kaiser et al. 2008), the Parker Solar Probe (PSP; Fox et al. 2016), and the Solar Orbiter (SolO; Müller et al. 2020). The last three spacecraft STEREO-A, PSP and SolO were separated from the Earth-Sun line by about 37.5°, 54°, 4° and located at distances of 0.96, 0.60, and 0.80 AU, respectively.

Various multi-wavelength imaging data unambiguously indicate that the October 28 SEP–GLE73 was initiated by the eruptive X1.0/2N-class flare and the associated halo CME. The flare with a soft X-ray peak at 15:35 UT occurred in the central-southern active region AR12887 at coordinates S26W05. It was a long-duration flare with a classic two-ribbon geometry and loops connecting the ribbons (Battaglia, Collier, & Krucker 2022). An outstanding global extreme-ultraviolet (EUV) coronal wave with a circular propagating bright front and its chromospheric counterpart Moreton wave were recorded after this flare (Hou et al. 2022). The totality of observations made it possible to conclude that the EUV wave front, driven by the expansion of the associated CME, had a three-dimensional dome-like shape propagating forward inclined to the solar surface. The estimated plane-of-sky speed of CME and its nose white-light shock were about 1240 and 1640 km $s^{-1}$, respectively, i.e. quite high, but noticeably less than in most GLEs (see Gopalswamy et al. 2012; Li et al. 2022; Papaioannou et al. 2022). The intensive type II and III radio bursts with components from metric to kilometer wavelengths accompanied the flare and CME (Klein et al. 2022).

Papaioannou et al. (2022) and Mishev et al. (2022) carried out a detailed analysis of the GLE73 characteristics based on both NM and space-borne data. The maximum enhancement of the count-rate of the relativistic proton flux up to 7.3% with a moderately hard rigidity spectrum was detected at two high-altitude polar conventional (i.e. lead) NM stations. The greatest count rate up to 14% was recorded by the same stations but with the bare (lead-free) NMs. The GLE flux remained above the background level for almost 4.5 hours. It was found that the particle flux arrived from the sunward direction and hence GLE73 was characterized by a moderate anisotropy.

Approximately at the time of release of relativistic protons from the Sun, the CME and associated white-light-driven shock were located at a height of (1.5–1.8)Rs and ~2.3Rs above the chromosphere, respectively. According to Papaioannou et al. (2022), the estimated time of the GLE protons release seems to be in good agreement with the time of EUV wave evolution towards the field lines magnetically connected to the Earth near AR12886 (S19W66). Klein et al. (2022) explored the NM proton data, space-borne measurements of relativistic electrons, dynamic spectra of decametric-to-kilometric type III radio bursts, as well as the in situ detection of the associated Langmuir waves near the Earth and proposed an alternative scenario. They conjectured that the high energy particles were accelerated in the lower corona during the eruption process, confined in the expanding magnetic structures and, then, got access to the Earth due to reconnection with open field lines.

As for the SEP characteristics, the peak proton flux with E >10 MeV turned out only ~30 pfu (1 pfu = 1 particle cm$^{-2}$ s$^{-1}$ sr$^{-1}$) which is lower than that observed in most GLEs (Papaioannou et al. 2022; Li et al. 2022). The fact that the >10 MeV proton flux remained increased (albeit with a softer spectrum) for almost a week was due primarily to the delayed prolonged acceleration, associated with the post-eruption energy release (see below) and the CME-driven shock from the given flare, and, then, to a number of additional overlapping proton flares and CMEs. Besides that, a contribution to this prolonged flux was also made by the low-energy, so-called energetic storm particles (ESPs; e.g. Lee, Mewaldt, & Giacalone 2012; Chiappetta et al. 2021), connected with the arrival to the Earth of ICMEs and the associated shock waves coinciding with geomagnetic storms.

## 3 PARAMETERS OF SEPS

The analysis of GLEs, known for example, from Logachev's catalogs of the solar proton events (see Logachev (2022) and references therein to the previous catalogs; http://www.wdcb.ru/stp/data/SPE/), shows that a sudden burst of the relativistic proton flux is recorded by NMs as a GLE, caused by the cascades of secondary particles in the Earth's atmosphere, when the intensity of the >1 GeV particles exceeds $(1–2)\times10^{-2}$ pfu (see Poluianov et al. 2017). This condition is easily satisfied for events with a sufficiently large flux of the >10 and >100 MeV protons, even though the integrated energy spectrum often steepens (softens) quite sharply when going from SEP to GLE. If the SEP flux does not exceed several tens of pfu, the GLE level can be reached only with a very flat (hard) SEP spectrum. The SEP–GLE73 of 2021 October 28 is just such an event.

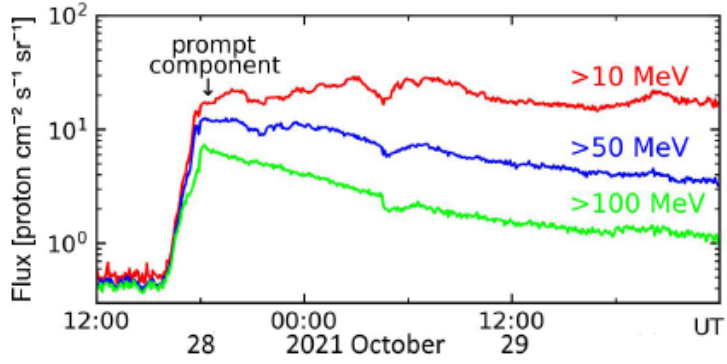

**Figure 1.** Time profiles of the GOES-16 proton fluxes with $E > 10$, 50, and 100 MeV. The prompt SEP component is marked with a vertical arrow.

From the SEP time profiles recorded by GOES-16 and presented in Fig. 1, it is evident that the prompt component of this event is characterized not only by a low >10 MeV proton flux (~20 pfu), but also by a notably hard energy spectrum. The fact that the indicated >10 Mev flux is indeed very small becomes apparent when compared with the mean and median values for 46 GLEs observed since 1976 and given by Papaioannou et al. (2022) and amount to 1850 and 321 pfu, respectively. The hardness of the SEP spectrum immediately follows from the uncommon quantitative proximity of the peak fluxes at >10 MeV and >100 MeV seen in Fig. 1. Observations on board of the SOHO, SolO, Wind, and STEREO-A spacecraft presented in Li et al. (2022) and Papaioannou et al. (2022) also confirm a very hard spectrum of the given SEP in the range of 10–100 MeV.

Let us analyze other SEP–GLE events in terms of ratio and spectrum of the peak flux of the >10 Mev ($J_{10}$) and >100 MeV ($J_{100}$) protons. For this analysis, we have taken GLEs associated with flares in the western half of the visible solar disk that occurred during Cycles 22–25, i.e. from 1987 to present time, and have considered the arrangement of events on the $J_{10}$–$J_{100}$ plane. The events associated with western beyond-limb flares have been excluded from consideration because occultation of their microwave radio emission, which will be analyzed in the next section. In cycle 21 (1977–1985), when regular SEP observations began with the GOES spacecraft, there were no GLEs with low $J_{10}$ flux.

Based on the GLE list (e.g. http://cosrays.izmiran.ru), we have formed a set of 26 events. The basic information on SEPs with the corresponding GLE numbers is provided in Table 1. It includes information on the flare that was the source of each event, the $J_{10}$, $J_{100}$ magnitudes, and the logarithm of their ratio $\gamma = log(J_{10}/J_{100})$. The latter can be conditionally regarded as an index of the power-low integral energy spectrum of protons in the range of 10–100 MeV. The SEP parameters for the prompt component were determined according to the GOES SEP plots similar to Fig. 1 (ftp://ftp.swpc.noaa.gov/pub/warehouse/) and data from catalogs edited by Logachev (2022). In Fig. 2, in addition to the distribution of events on the $J_{10}$–$J_{100}$ plane, straight lines are drawn corresponding to the power-low spectral indexes $\gamma$ =0.7, 1.0, and 1.5. This original presentation of the data proved to be very informative.

**Table 1.** Parameters of the prompt component of the >10 MeV and >100 MeV proton fluxes in GLEs under consideration.

| GLE No. | Date | Flare coordinates | Proton flux $J_{10}$ (pfu) | Proton flux $J_{100}$ (pfu) | Spectral index $\gamma$ |
|---|---|---|---|---|---|
| 40 | 1989.07.25 | N26W85 | 27 | 5,2 | 0.71 |
| 41 | 1989.08.15 | S16W83 | 1000 | 42 | 1.38 |
| 44 | 1989.10.22 | S27W32 | 2500 | 140 | 1.25 |
| 45 | 1989.10.24 | S29W57 | 2200 | 120 | 1.26 |
| 46 | 1989.11.15 | N11W2 | 25 | 2.7 | 0.97 |
| 47 | 1990.05.21 | N34W37 | 160 | 11 | 1.16 |
| 48 | 1990.05.24 | N36W76 | 180 | 11 | 1.21 |
| 51 | 1991.06.11 | N32W15 | 300 | 12 | 1.4 |
| 52 | 1991.06.15 | N36W70 | 750 | 45 | 1.22 |
| 53 | 1992.06.25 | N09W69 | 130 | 7.5 | 1.24 |
| 55 | 1997.11.06 | S18W63 | 190 | 49 | 0.59 |
| 56 | 1998.05.02 | S15W15 | 130 | 6.8 | 1.27 |
| 57 | 1998.05.06 | S15W64 | 120 | 2.5 | 1.68 |
| 59 | 2000.07.14 | N22W07 | 2200 | 330 | 0.82 |
| 60 | 2001.04.15 | S20W84 | 380 | 138 | 0.44 |
| 62 | 2001.11.04 | N07W19 | 1100 | 59 | 1.27 |
| 63 | 2001.12.26 | N08W54 | 520 | 47 | 1.17 |
| 64 | 2002.08.24 | S02W81 | 220 | 26 | 0.91 |
| 66 | 2003.10.29 | S15W02 | 2100 | 105 | 1.30 |
| 67 | 2003.11.02 | S14W56 | 800 | 42 | 1.28 |
| 68 | 2005.01.17 | N14W24 | 1800 | 26 | 1.84 |
| 69 | 2005.01.20 | N12W58 | 1600 | 680 | 0.37 |
| 70 | 2006.12.13 | S06W24 | 660 | 86 | 0.89 |
| 71 | 2012.05.17 | N11W76 | 230 | 18 | 1.11 |
| 72 | 2017.09.10 | S11W90 | 1000 | 40 | 1.4 |
| 73 | 2021.10.28 | S26W05 | 20 | 6.6 | 0.48 |

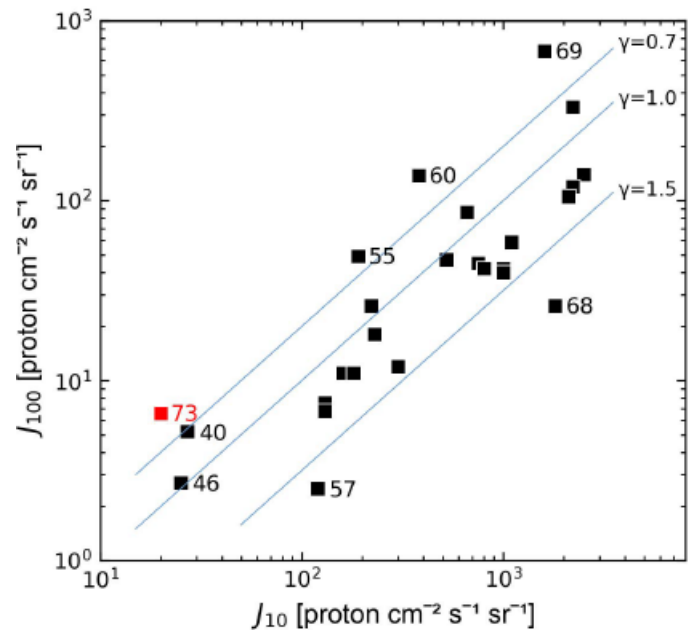

**Figure 2**. Scatter of the western GLEs observed in Cycles 22–25 on the $J_{10}$–$J_{100}$ proton flux plane. Numbers indicate the generally accepted GLE numbers (see Table 1). Red square belongs to SEP–GLE73. The blue lines correspond to the specified power-low spectral indexes $\gamma$ in the energy range of 10–100 MeV.

As seen in Fig. 2, GLE73 (red square) really differs from most of the other GLEs under consideration both by a low SEP flux ($J_{10}$~20, $J_{100}$~6.6 pfu) and its hard energy spectrum ($\gamma$~0.48). Only two events GLE40 (1989 July 25) and GLE46 (1989 November 15) have comparable SEP parameters and can be considered as close analogues of GLE73. Recall that GLE73 occurred at the very beginning of the ascending phase of Cycle 25 from the flare slightly west of the central meridian. In contrast to this, GLE40 and GLE46 occurred near the maximum of Cycle 22 and were associated with flares located much more to the west (see Table 1). It is noteworthy that in all these three events, the >100 MeV proton flux was confined to the range $J_{100}$~(2.7–6.6) pfu. In Fig. 2, there are three other events with similar $J_{100}$ magnitudes, but their $J_{10}$ flux is about 130 pfu, which corresponds to their softer SEP spectrum with $\gamma$~1.3–1.7. Note that the analogs GLE40 and GLE46 displayed approximately the same maximum increase of the relativistic proton flux on conventional NMs (about 8% and 12%) as GLE 73 (7.3%) (Belov et al. 2010).

## 4 SEPS AND RADIO SPECTRA

It is known that the intensity and frequency spectra of the 3–15 GHz microwave bursts, though they are generated by electrons propagating towards the photosphere, reflect the number and the energy spectrum of accelerated particles, including protons arriving at the Earth with an energy of tens of MeV (see Klein (2021a,b) for a review). In particular, a statistical correspondence is established for the western events between the $\gamma$ index of the SEP energy spectrum and the microwave burst parameters such as the frequency of the spectral maximum $f_m$ and the ratio of the peak radio flux densities at a pair of the highest frequencies (Chertok 1982, 1990; Chertok et al. 2009). It turned out that SEPs with a hard proton spectrum tend to be associated with strong microwave bursts, whose radio spectrum noticeably grows (hardens) to high frequencies.

It is reasonable to look at GLE 73 from this point of view. Unfortunately, for GLE73, we only have patrol observation data of the USAF Radio Solar Telescope Network (RSTN; https://www.sws.bom.gov.au/World_Data_Centre/2/8/9) from the Sagamore Hill Radio observatory at meter and decimeter wavelengths, but no information is available about the microwave bursts (ftp://ftp.swpc. noaa.gov/pub/warehouse/). Therefore, it only remains to analyze the radio characteristics of the analogous GLE40 and GLE46, as well as of other GLEs from our selection using data provided by the four RSTN patrol stations.

Fig. 3a shows that both GLE40 and GLE46 (which have a hard SEP energy spectra as demonstrated in Section 3) also display a hard radio spectrum with increasing microwave flux density up to the highest observed frequency 15.4 GHz. Moreover, a similar hard radio spectrum is typical of the other three GLEs with a hard SEP spectrum of $\gamma$ <0.7 (see Fig. 2). In contrast, the GLE57 and GLE 68 with the softest SEP spectra $\gamma$ >1.5, display accordingly much softer radio spectra with the pick flux densities at relatively low frequencies $f_m$~ 8.8 and 5 GHz (Fig. 3b).

A comparison of the radio spectra shown in Fig. 3 with the arrangement of events in Fig. 2 demonstrates that microwave bursts reflect not only the energy spectra of SEPs, but also the intensity of proton fluxes especially in the energy range of 10–100 MeV: the larger microwave bursts correspond to more intense SEPs (although within the pair of the similar events GLE40 and GLE46 this trend is reversed). For example, the well-known events of 2005 January 17 and 20 (GLE68 and GLE69) have approximately the same proton flux $J_{10}$, but, as illustrated in Fig. 2, they strongly differ in the $J_{100}$ flux. It corresponds to that (as comparison of panels *a* and *b* of Fig. 3 displays) these GLEs have the relatively soft (hard) microwave radio and soft (hard) SEP energy spectra, respectively.

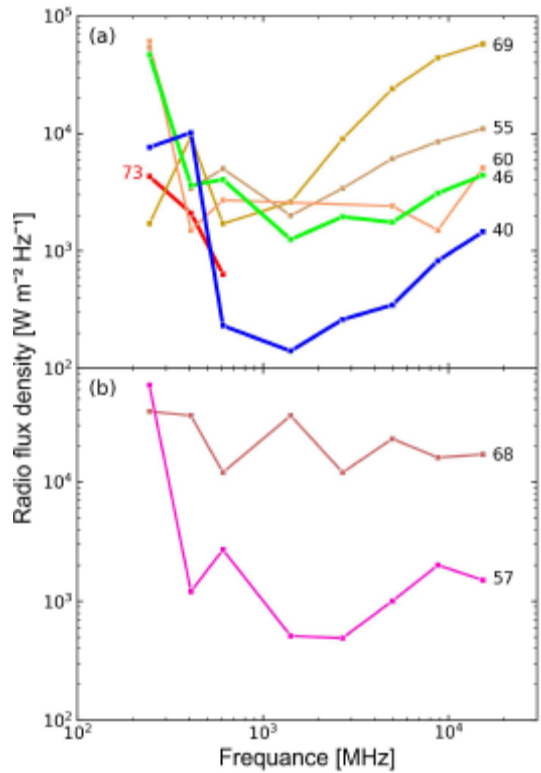

**Figure 3.** Frequency spectra of radio bursts of the GLE events with the hard (*panel a*) and relatively soft (*b*) SEP energy spectra in the 10–100 MeV range. The indicated numbers correspond to the GLE numbers presented in Fig. 2 and Table 1. The incomplete available data of GLE73 are shown on *panel a*.

All these regularities give grounds to suggest that SEP-GLE73, we are interested in, most likely could have a hard microwave radio spectrum similar by the character and flux density to something in between those of GLE40 and GLE46. However, the situation is not so clear. In their recent paper, Klein et al. (2022) presented the 1–18 GHz microwave spectrum of the X1.0-class flare, which initiated the SEP-GLE73 event, as observed with the Extended Owens Valley Solar Array (EOVSA; Gary et al. 2018). According to these data, the spectrum also had a low maximum flux density of about 1000 sfu, but peaked at a frequency near 5 GHz, i.e. was rather soft than hard. However, the evolution of the EOVSA spectrum indicates to its continued hardening throughout the rise and post-peak phases of the microwave burst. If these EOVSA data are confirmed, we will have to consider the SEP-GLE73 event as an exception by the character of the radio spectrum. Nevertheless this does not change the conclusion about general correspondence between the flare-generated microwave spectra of the peak radio flux density, on the one hand, and the intensity and energy spectra of the SEP–GLE events, on the other.

**5 SUMMARY AND DISCUSSION**

The paper by Papaioannou et al. (2022) is mainly devoted to a detailed analysis of the GLE73 relativistic proton flux by the NM data, as well as to consideration of the initiating flare and CME. In addition to this work and the papers by Klein et al. (2022) and Mishev et al. (2022), we have analyzed the 10–100 MeV SEP component of this GLE using the GOES satellite and radio network observations. The study was carried out by consideration of the arrangement of SEP–GLE73 and other 26 GLEs at the $J_{10}$–$J_{100}$ plane with designation of the conditional power-law energy spectral indices, as well as by comparison of the SEP energy and microwave radio spectra. The results can be summarized as follows:

(**i**) The SEP-GLE73 prompt component fluxes are confirmed to be among the lowest for both $J_{10}$~20 pfu and $J_{100}$~6.6 pfu.

(**ii**) The unusual quantitative proximity of these peak fluxes means that the integral energy spectrum of the given SEP component was extremely hard with a power-law index $\gamma$~0.48.

(**iii**) According to the above parameters, only two events GLE40 (1989 July 25) and GLE46 (1989 November 15) are analogues of SEP–GLE73.

(**iv**) These two GLE events and three other GLEs with a hard SEP spectrum are associated with flares that have hard radio spectra in the microwave range with the flux density increasing towards high frequencies, while GLEs with a relatively soft SEP spectrum are associated with flares displaying a soft radio spectrum.

Noticeable SEP–GLE events are usually observed after major solar eruptions including both strong long-duration flares and fast, wide CMEs. In this regard, one must keep in mind at least three sources of particle acceleration (see reviews in the Topical issue, edited by Gopalswamy &, Nitta (2012), as well as references in the informative introductions to Letter by Papaioannou et al. (2022) and to the recent papers by Kiselev, Meshalkina, & Grechnev (2022), and Klein et al. (2022)). The first acceleration source is identified with an impulsive (primary) phase of the flare as a result of the magnetic reconnection by interaction of opposite polarity loops in the upper chromospheres and lower corona and contributes to the prompt SEP–GLE component (see McCracken, Moraal, & Shea 2012).

The two other acceleration sources are closely associated with CMEs. One of them takes place when the AR magnetosphere strongly disturbed by a large CME, relaxes to reach a new quasy-equilibrium state via the magnetic reconnection in a vertical current sheet high in the corona. It is accompanied by a prolonged post-eruption (PE) energy release, an effective particle acceleration, generation of long duration delayed emissions in the X-ray, gamma-ray, microwave, and other ranges, the formation of two-ribbon flare structure, post-flare loops, giant arches, etc. (e.g. Svestka 1989; Chertok 1995). These processes, which are typical of eruptive flares, are treated usually as a standard model of solar flares.

Almost simultaneously with the PE acceleration, also high in the corona in front of the fast rising CME, a piston shock wave is formed, which is the third source of particle acceleration in eruptive flares (e.g. Cliver et al. 2004; Gopalswamy et al. 2012; Kahler et al. 2017). Herein, protons accelerated during the impulsive and PE flare phases can serve as a seed population for acceleration in the CME-driven shock. From the picture outlined above, it is clear that the contributions and manifestations of the PE particle acceleration and acceleration in the CME-associated shock are difficult to unambiguously identify and distinguish from each other (see Aschwanden 2012; Reames 2020).

The results of our analysis of the SEP–GLE73 and its analogs in terms of the intensity and energy spectrum of proton fluxes in the range of 10–100 MeV, as well as of the relationship of these characteristics with the flux density and frequency spectrum of the corresponding microwave radio bursts, indicate a significant role of the flare (both impulsive and PE phases) in the acceleration of the SEP-GLE protons. This does not rule out a possible additional particle acceleration in the CME-driven shock. The statement about the flare origin of the SEP-GLE particles is consistent with conclusions of Mishev et al. (2022) and Klein et al. (2022) that the prompt component of SEP-GLE73 event was accelerated low in the corona and released due the trap-plus-reconnection processes.

## 6 ACKNOWLEDGEMENTS

The author thanks the anonymous reviewer for useful remarks and comments. The author is grateful to the GOES, RSTN, NMDB and other teams for open access to their data. Also thanks to A.A.Abunin, A.V. Belov, and A. Papaioannou for helpful discussions and suggestions.

## DATA AVAILABILITY

The sample of the analyzed events was formed based on the GLE list (e.g. http://cosrays.izmiran.ru). The SEP data and plots used in this study are available at the GOES site ftp://ftp.swpc.noaa.gov/pub/warehouse/. The Catalogs of the solar proton events for solar cycles 20–24, edited by Yu.I. Logachev, are accessible at http://www.wdcb.ru/stp/data/SPE/. The parameters of the corresponding microwave radio bursts were determined from data of the USAF Radio Solar Telescope Network (RSTN; https://www.sws.bom.gov.au/World_Data_Centre/2/8/9).